\documentclass[final,twocolumn,times,floatfix]{revtex4}

\usepackage{amssymb}
\usepackage{amsmath}
\usepackage{textcomp}
\usepackage[dvips]{graphicx}

\begin{document}

\title{Energy filtering enhancement of  thermoelectric performance of nanocrystalline Cr$_{1-x}$Si$_x$ composites}

\author{A. T. Burkov }
\author{S. V. Novikov}
\affiliation{A.F. Ioffe Physical-Technical Institute, Saint Petersburg, 194021, Russia\\
and ITMO University, Saint-Petersburg, 197101, Russia.}
\author{V. V. Khovaylo}
\affiliation{Department of Functional Nanosystems and High-temperature Materials, National University of Science and Technology ‘‘MISiS’’, Moscow,  119049, Russia}
\author{J. Schumann}
\affiliation{Leibniz Institute for Solid State and Materials Research, Dresden, Germany.}
\date{\today}

\begin{abstract}

We report on thermoelectric properties of nanocrystalline Cr$_{\rm 1-x}$Si$_{\rm x}$ composite films. As-deposited amorphous films were  transformed into a nanocrystalline state with average grain size of 10--20~nm by annealing during in-situ thermopower and electrical resistivity measurements.  
The partially crystallized films, i.e. the films consisting of crystalline grains dispersed in the amorphous matrix, are a new type of the heterogeneous material where the nanocrystalline phase plays the role of scattering centers giving rise to a large contribution to the thermopower.
We show that the thermopower enhancement is related to the energy dependent scattering (energy filtering) of the charge carriers on the nanograin interfaces.

\end{abstract}
\maketitle

Nanocrystalline composite, thermoelectric properties, electronic transport 



\section{Introduction}
Nanocrystallization, or - more generally, nanostrucrturing has been considered as a promising way for essential improvement of conversion efficiency of thermoelectric devices \cite{rowe2006, kanatzidis2010}.
There are several approaches for preparing nanocrystalline (NC) thermoelectric compounds.
Among them, crystallization from the amorphous state features low level of contamination of NC composite by environmental impurities, which may have a decisive impact on thermoelectric performance of NC material.
In our recent article \cite{novikov2013} we have demonstrated that NC CrSi$_{\rm 2}$ and MnSi$_{\rm 2}$ compounds, prepared by crystallization from the amorphous state, have a higher thermoelectric power factor $P= S^{\rm 2}\sigma$  ( $S$ is thermopower, or Seebeck coefficient, $\sigma $ is electrical conductivity) in comparison with the polycrystalline counterparts.
This result implies a considerable increase of thermoelectric performance in NC materials.
Understanding mechanisms underlying nanocrystallization process and transport properties in different structural  states is of a crucial importance for the development of practically useful nanostructured thermoelectrics.
It has been suggested that the observed in \cite{novikov2013} power factor enhancement arising due to increasing thermopower in NC composites can be connected to the energy filtering effect due to scattering at nanograin interfaces (see also \cite{burkov1996b,burkov98b}).

It has been shown theoretically that scattering on metal/semiconductor interfaces in layered structures or on metallic nanoinclusions in a semiconducting host can lead to a significant increase of thermopower due to the energy filtering effect \cite{moyzhes1998,faleev2008}. 
Later on this idea has been utilized for the interpretation of experimental results obtained for nanocomposites based on Sb$_{\rm 2}$Te$_{\rm 3}$ matrix with nanoinclusions of Pt \cite{ko2011} or  $\gamma$-SbTe \cite{kim2015a}, Bi$_2$Te$_3$ thin films with native defects \cite{suh2015},  nanocrystalline silicon with inclusion of a secondary phase \cite{narducci2015}, and nano-porous silicon \cite{zhang2015b}.

In this article we report on experimental investigation of conduction mechanism  in the nanocrystalline Cr$_{\rm 1-x}$Si$_{\rm x}$ ($0.65 < x < 90$) composites, particularly on the investigation of the contribution of the charge carrier scattering at nanograin interfaces to thermopower and   electrical resistivity.
We give a direct experimental evidence that scattering of the charge carriers at nanograins of the semiconducting CrSi$_{\rm 2}$ diluted in an amorphous metallic-like Cr-Si matrix gives rise to a strong increase of thermopower due to the energy filtering effect.

\section{EXPERIMENTAL PROCEDURES}

Amorphous films were prepared by magnetron sputtering from composite targets onto unheated substrates. 
The films thickness was about 100 nm. Si wafers, thermally oxidized to form a 1 $\mu $m layer of SiO$_{2}$, were used as the substrates. 
The  SiO$_{2}$ layer provides electrical insulation of the films from Si substrate and prevents chemical reactions of the films with the substrate during annealing and transport property measurement procedures. 
The film composition was determined with Rutherford Backscattering Spectroscopy (RBS) and Energy Dispersive X-ray Spectroscopy (EDX).
The composite films were produced by crystallization from the amorphous state in the course of annealing of the films in a high-purity helium atmosphere during in-situ transport measurements. 
The state of the film composite in the course of its transformation from amorphous to NC composite was controlled by means of the in-situ transport measurements -- simultaneous measurements of electrical resistivity ($\rho = 1/\sigma $) and $S$. 
Standard DC 4 points configuration was used in the resistivity measurements, while differential method was utilized for thermopower \cite{burkov2001}.  Transmission Electron Microscopy and X-Ray Diffraction were utilized for structural characterization of the samples \cite{pitschke2001}.

\section{Experimental results and discussion}

\begin{figure}
 \begin{center}
 \includegraphics[scale=0.45]{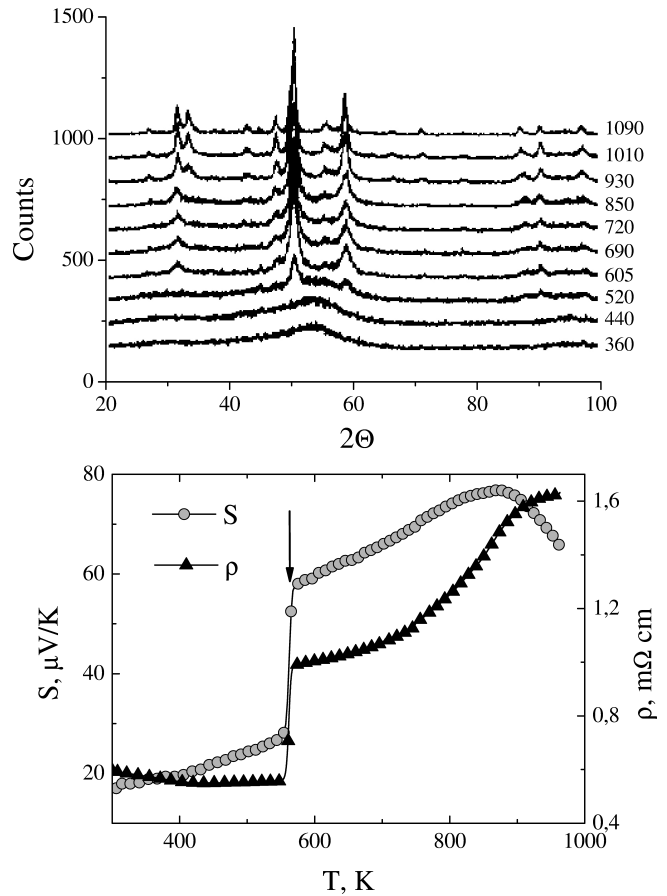}
\end{center}

\caption{Upper panel shows in-situ XRD spectra of  Cr$_{\rm 0.28}$Si$_{\rm 0.72}$ thin film composite at different annealing stages. Annealing temperature in Kelvin is indicated next to each XRD scan. At each scan, the sample was annealed for about 30 min. The lower panel presents temperature dependences of electrical resistivity ($\rho$) and thermopower ($S$) of Cr$_{\rm 0.28}$Si$_{\rm 0.72}$ film measured at a constant heating rate of 5 K/min. \label{x-ray}}
\end{figure}

\begin{figure}
 \includegraphics[scale=0.5]{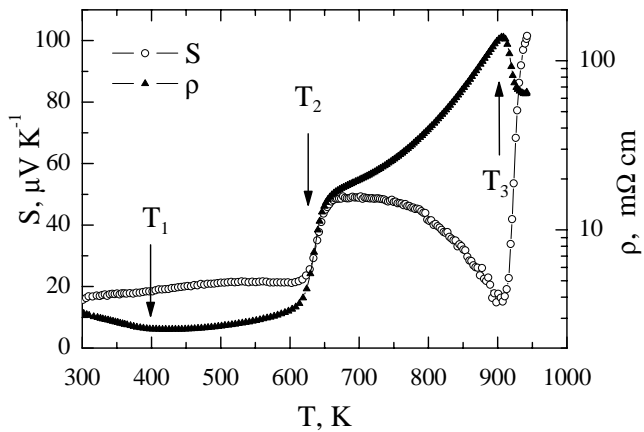}
\caption{Temperature dependences of electrical resistivity $\rho (T)$ and thermopower $S(T)$ of Cr$_{\rm 0.13}$Si$_{\rm 0.87}$ film. T$_{\rm 1}$, T$_{\rm 2}$ and T$_{\rm 3}$ indicate temperature of structural relaxation in the amorphous state, crystallization temperature, and temperature of formation of percolating cluster of NC phase, respectively. \label{r&s-T}}
\end{figure}

\begin{figure}
 \begin{center}
 \includegraphics[scale=0.3]{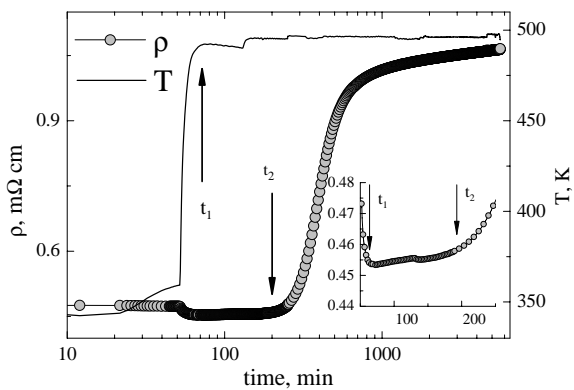}
\end{center}
\caption{Resistivity of Cr$_{\rm 0.33}$Si$_{\rm 0.67}$ film composite and temperature of the sample as functions of time. The inset shows the resistivity dependence during incubation period $t _{\rm inc}=t_{\rm 2}-t_{\rm 1}$ at extended scale. \label{r-T-time}}
\end{figure}

According to X-ray data (Fig.~\ref{x-ray}, upper panel), the as-deposited films are amorphous.
The X-ray data indicate that the onset of crystallization takes place after annealing above 500~K \cite{schumann94}.
The crystallization has a dramatic impact on electronic transport (see Fig~\ref{x-ray}, lower panel), therefore the transport properties can be used as sensitive probes to detect and monitor structural transformations in the films.
As another example of such effect, Figure~\ref{r&s-T} presents temperature dependences of thermopower and electrical resistivity of a Cr$_{\rm 0.13}$Si$_{\rm 0.87}$ film.
There are well defined temperatures at which both properties undergo sudden changes as temperature increases.
Comparing with x-ray data one can conclude that the sharp increase of both, $\rho $ and $S$ at T$_{\rm 2}$ reflects the onset of the crystallization.
It has been shown that amorphous Cr$_{\rm 1-x}$Si$_{\rm x}$ films crystallize into nanocrystalline composite with average grain size about 10--20~nm, this state is stable to about 1000~K \cite{gladun94}.
When comparing the XRD data (Fig.~\ref{x-ray})  and temperature-dependent transport properties results (Figs.~\ref{x-ray}, \ref{r&s-T}), one has to remember that, while the XRD experiment involves about 30 min annealing of the sample at the temperature of each XRD scan, the transport measurements were made with continuously varying temperature with a rate of about 5~K/min.
Since the crystallization is characterized by an incubation time, $t_{\rm inc}$ (see Fig.~\ref{r-T-time}), which is strongly dependent on annealing temperature \cite{pitschke2001},  the peculiarity in the $\rho (T)$ and $S(T)$ dependences are observed at a higher temperature in comparison with XRD.

 
Another clear peculiarity at T$_{\rm 3}$ is observable in films with excess of silicon in comparison with CrSi$_{\rm 2}$ composition. 
As it was shown earlier for Re-Si  and for Cr-Si film composites \cite{burkov98b,burkov2004c}, this reflects formation of percolation cluster of NC stoichiometric chromium disilicide.

One can make the in-situ annealing in two modes: 
\begin{enumerate}
 \item 
As temperature-dependent measurements of $\rho (T)$ and $S(T)$, as is shown in Fig.~\ref{r&s-T}; 
\item
As isothermal  annealing, with measurement of $\rho (t)$ and $S(t)$ as a function of annealing time ($t$), Fig.~\ref{r-T-time}.
\end{enumerate}
In the temperature-dependent measurements the experimental $\rho(T)$ and $S(T)$ result from a combination of two principle contributions:
\begin{itemize}
 \item 
Intrinsic temperature dependence of the material properties.
\item 
Modifications of the intrinsic temperature dependences due to structural transformations in  the course of the annealing.
These modifications develop in time with the rate which depends on temperature.
\end{itemize}

Isothermal annealing apparently gives direct information about $\rho (t)$ and $S(t)$ dependences due to structural modifications.
However, this is true only for structural transformations with comparatively long incubation time $t _{\rm inc}$ (for terminology see, for example, \cite{ruitenberg2002}), i.e., the time span between the moment when sample temperature reaches the target annealing temperature and the apparent beginning of transformation (nucleation stage followed by grain growth).
An example of such situation is shown in Fig.~\ref{r-T-time}, where temperature profile and resistivity of a Cr$_{\rm 0.33}$Si$_{\rm 0.67}$ film composite are given as a function of annealing time. 
The sample temperature was stabilized at $t=t_{\rm 1}$. 
At earlier times, $t < t_{\rm 1}$, the resistivity variation with time results from increase of the sample temperature and is governed by the  intrinsic temperature dependence of the resistivity $\rho (T)$. 
It is seen from Fig.~\ref{r-T-time} that the resistivity is almost constant during some period of time, from $t_{\rm 1}$ to $t_{\rm 2}$, and starts to increase drastically at $t > t_{\rm 2}$.
Since at $t>t_{\rm 1}$ temperature is stable, the observed resistivity variation at $t>t_{\rm 2}$ is related only to annealing effects.
The time interval $t_{\rm inc}=t_{\rm 2}-t_{\rm 1}$ can be regarded as an incubation time. 
When the incubation time is short on the time-scale of annealing temperature stabilization, $\rho (t)$ and $S(t)$ dependences at the beginning of transformation in isothermal annealing procedure will be distorted by the combining effects of temperature variation and the structural transformation. 

\begin{figure}[htb]
 \begin{center}
 \includegraphics[scale=1]{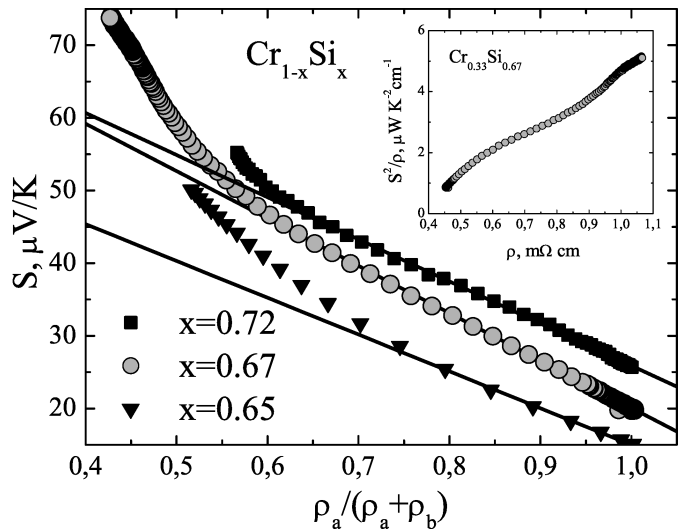}
\end{center}
\caption{The Nordheim-Gorter plot for Cr$_{1-x}$Si$_{x}$  films ($ x=$ 0.72, 0.67 and 0.65). The solid straight lines show the Nordheim-Gorter relation (2), fitted to the experimental data. The inset presents dependence of the power factor  $S^{\rm 2}/\rho$ of Cr$_{\rm 0.33}$Si$_{\rm 0.67}$ composite against resistivity. \label{nordg-1}}
\end{figure}

It has been shown that at temperatures well below the temperature of the formation of percolating cluster of the crystalline phase T$_{\rm 3}$, the NC grains do not contribute directly to conduction process, being isolated within the amorphous matrix \cite{burkov98b,burkov2004c}.
Between T$_{\rm 2}$ and T$_{\rm 3}$ the film represents a composite consisting of CrSi$_{\rm 2}$ nanograins dispersed in the amorphous Cr-Si matrix. 
At early crystallization stage, when the amount of the NC phase is small, the individual nanograins represent non-interacting scattering centers for the charge carriers.
To a first approximation, such composite can be considered as a binary diluted alloy of nanograins in the amorphous  Cr-Si matrix.
Accordingly, the well known Matthiessen and Nordheim-Gorter rules \cite{rossiter1987,barnard1972} can be applied for analysis of the electrical resistivity ($\rho$) and thermopower ($S$) of the composite:
\begin{equation}{}
 \rho = \rho _{\rm a} + \rho _{b}
\end{equation}
\begin{equation}{}
S = (S_{\rm a}-S_{\rm b})\frac{\rho _{\rm a}}{\rho _{\rm a}+\rho _{\rm b}} + S_{\rm b}, 
\end{equation}
where $\rho _{\rm a}$ and $S_{\rm a}$ are  resistivity and thermopower of ``pure'' amorphous matrix, $\rho _{\rm b}$ and $S_{\rm b}$ are the contributions to the resistivity and thermopower of the matrix due to the scattering of charge carriers at nanograin boundaries.
If the made above assumption  is valid, the plot of total thermopower of the partially crystallized composite against reciprocal total resistivity should be a straight line. 
Fig.~\ref{nordg-1} shows such plot for Cr$_{1-x}$Si$_{x}$ films measured during isothermal annealing (see Fig~\ref{r-T-time}).

\begin{figure}[htb]
 \begin{center}
 \includegraphics[scale=0.35]{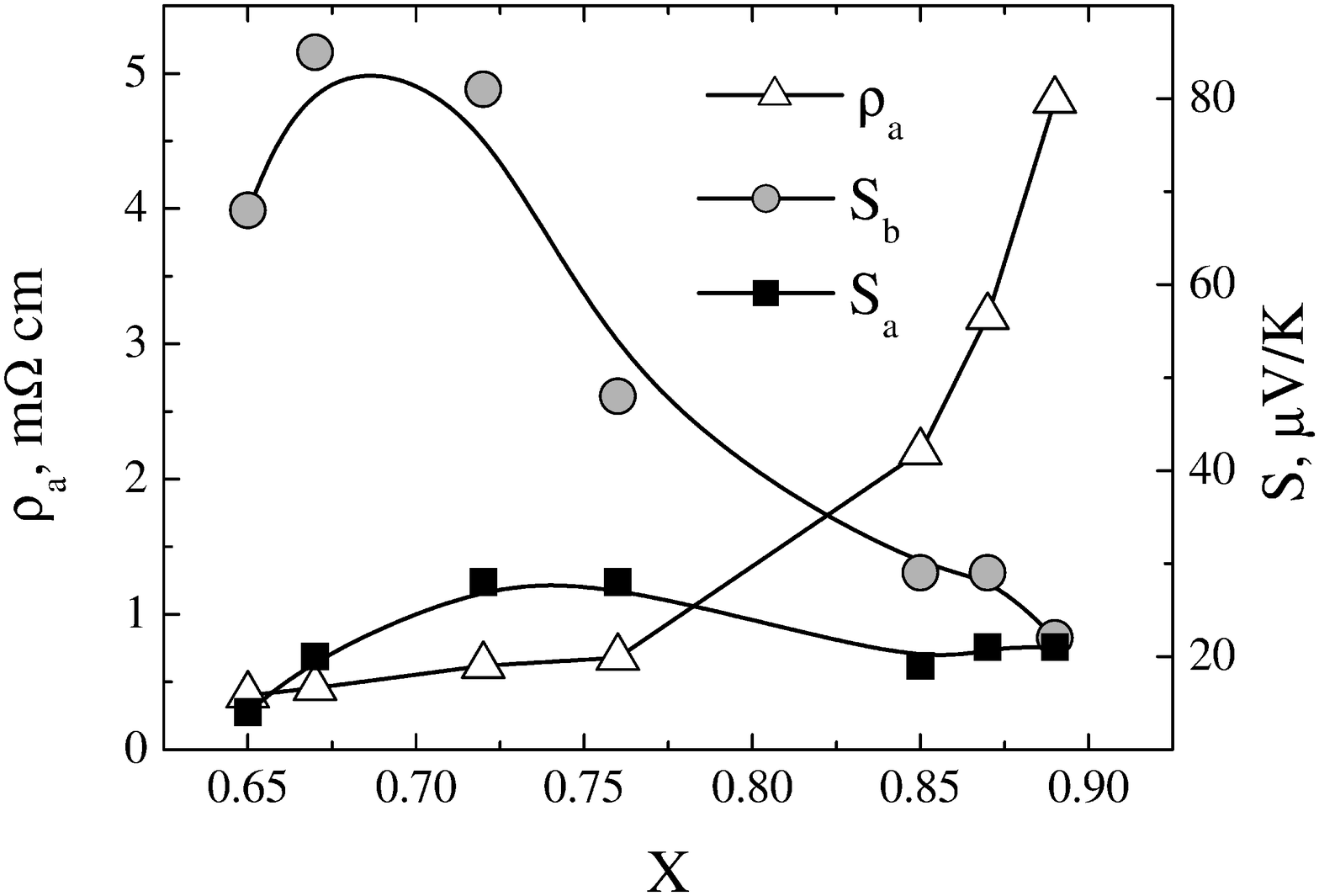}
\end{center}
\caption{Grain boundary contribution to thermopower and resistivity as a function of the film composition. \label{s-r-comp}}
\end{figure}

We observed a good linear dependence in a broad range of the variation of $\rho = \rho _{\rm a} + \rho _{\rm b}$.
From the plot one can determine the thermopower of ``pure'' amorphous matrix $S_{\rm a}$ and the grain boundary contribution $S_{\rm b}$.
Applying this procedure to the films of different compositions, the dependence of $S_{\rm a}$, $S_{\rm b}$ and $\rho _{\rm a}$ on the Si/Cr ratio was determined and is presented in Fig.~\ref{s-r-comp}.
Two important points is necessary to emphasize: $(i)$ the grain boundary contribution has the largest value for the films of nearly stoichiometric CrSi$_{\rm 2}$ composition; $(ii)$ the grain boundary contribution  is large, in its maximum it is about 4 times larger than the intrinsic thermopower of the amorphous compound.
The latter observation means that the grain boundary scattering is highly selective, i.e., its scattering magnitude strongly depends on the charge carrier energy.
The resistivity of the composites also increases upon crystallization due to additional scattering on the interfaces.
However, the increase of thermopower outweighs the rise of the resistivity so that the power factor $S^{\rm 2}/\rho$ also increases in the NC composites, see inset in Fig.~\ref{nordg-1}.
The large magnitude of $S_{\rm b}$ and its strong dependence on the composition have important implications for the on-going development of the nanocrystalline thermoelectrics.
It is therefore important to understand the underlying physical mechanisms.
Clearly, properties of the interface between the amorphous matrix and the nanograin play an important role.

\begin{figure}
 \begin{center}
 \includegraphics[scale=0.45]{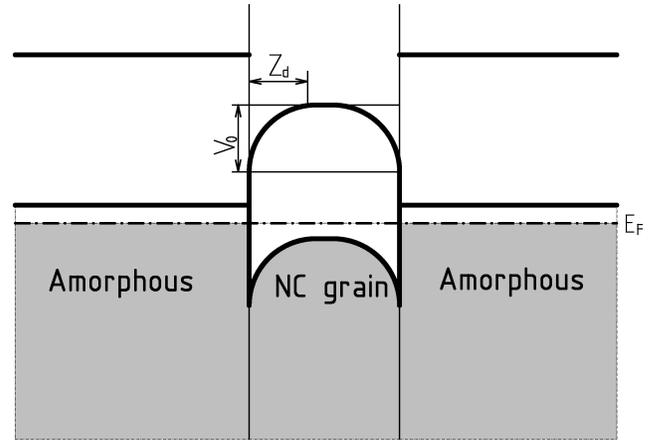}
\end{center}

\caption{Schematic energy diagram of the amorphous-nanocrystalline Cr-Si junction. \label{junc}}
\end{figure}

According to the behavior of the transport properties, the amorphous matrix can be described as an amorphous metal, whereas the crystalline phase is a p-type semiconductor with a band gap of about 0.4~eV \cite{novikov2014,vining95}.
Therefore, to a first approximation, one can model the interface as a metal-semiconductor junction.
The energy diagram of the model junction is shown in Fig.~\ref{junc}.

Due to the band bending at the interface there is a depletion region located mainly in the NC grains (due to a lower carrier concentration).
The depth of the depletion layer ($Z_{\rm d}$) can be estimated with expression \cite{kroezen2012}: $$Z_{\rm d}=\sqrt\frac{2\varepsilon \varepsilon _{\rm 0}V_{\rm 0}}{eN_{\rm p}},$$
where $\varepsilon $ is the relative permittivity of CrSi$_{\rm 2}$, $\varepsilon _{\rm 0}$ = 8.854 10$^{\rm -12}$~F/m is the permittivity of vacuum, $V_{\rm 0}$ is a magnitude of the band bending, $e$ is electron charge, and $N_{\rm p}$ is the charge (hole) carrier concentration.
The band bending magnitude $V_{\rm 0}$ is typically of order of half band gap $E_{\rm g}$.
Taking for the estimation the following parameters: $E_{\rm g} = $ 0.4~eV ($V_{\rm 0} \approx $ 0.2~eV), $\varepsilon \approx$ 30 \cite{bellani1992}, $N_{\rm p} \approx $ 10$^{\rm 24}$ - 10$^{\rm 25}$~m$^{\rm -3}$ \cite{nava1985}, one obtains $Z_{\rm d} $ in the range from 10~nm to 30~nm.
Considering that the average grain size in the partially crystallized film is of about 20~nm, this indicates that the whole volume of majority of nanograins will be strongly affected by the band bending effects at the interfaces.

Effectively, at moderate temperatures the nanograins represent the strongly scattering centers for holes: only high energy holes can penetrate the barrier at the amorphous-crystalline interface.
This energy-filtering mechanism results in a large positive contribution to the thermopower.
At the same time the contribution of the crystallized phase to the total conductivity is small in the dilute regime since the conduction in the nanograins is ballistic due to a small size of the grains. 
Rough estimate of the mean free path in crystalline CrSi$_{\rm 2}$ from electrical conductivity (assuming average hole velocity 10$^{\rm 5}$~m/s) yields $l \approx $ 10$^{\rm -7}$~m, i.e., about 100~nm, which is much larger than the mean size of the nanograins.
That means that the main scattering channel for the charge carriers in the nanograins is the scattering on the amorphous-crystalline interfaces.

An opposite situation of metallic nanoinclusions in a semiconducting host (PbTe) has been considered theoretically by Faleev and Leonard \cite{faleev2008}. 
They demonstrated that scattering of charge carriers on the metal/semiconductor interfaces significantly enhances thermopower due to energy filtering.
Similar mechanism was proposed for Pt nano-inclusions in Sb$_{\rm 2}$Te$_{\rm 3}$ nanocomposite \cite{ko2011}.

Obviously, there is an optimum effective barrier height resulting in the maximal  contribution to the thermopower \cite{ravich95}.
When the barrier height is large in comparison with the thermal energy of the charge carriers k$_{\rm B}$T (where k$_{\rm B}$ is Boltzmann constant), all carriers are scattered uniformly and there is no energy filtering and, as a consequence, no additional contribution to thermopower.
The upper limit for the thermal energy, set at about 1000~K by the band gap  of CrSi$_{\rm 2}$, is related to the onset of intrinsic conductivity.
On the other hand, for a very small barrier height, only a small fraction of the charge carriers will be filtered out by the scattering at the nanograin interfaces with corresponding small contribution to the thermopower.
The barrier height depends on the film composition: in stoichiometric films both NC and amorphous phases have the same chemical composition, therefore no segregation takes place during crystallization.
However, in non-stoichiometric films the excess element should diffuse to the grain boundary during crystallization.
Therefore, in films with excess of silicon the amorphous-crystalline interface should be enriched by silicon.
This shall increase the effective barrier height.
The effect of Cr excess is not so obvious, it can lead to a decrease as well as to an increase of the effective barrier height.
Nevertheless, the experimental results indicate (Fig.\ref{s-r-comp}) that optimum barrier height is realized in nearly stoichiometric films with small silicon excess.


\section{Conclusions}
The nanocrystalline Cr$_{\rm 1-x}$Si$_{\rm x}$ composites with the average grain size of 10--20 nm were prepared by crystallization of amorphous films  deposited by magnetron sputtering on unheated Si/SiO$_{\rm 2}$ substrates. 
 
Electrical resistivity and thermopower of the film composites at different stages of crystallization were measured from 100~K to 1000~K.
We show experimentally that the partially crystallized film, i.e., the films consisting of crystalline grains  dispersed in amorphous matrix, is a new type of the heterogeneous material, where the nanocrystalline phase plays role of the scattering centers giving rise to a large contribution to both thermopower and  electrical resistivity.
The increase of thermopower, however, outweighs the enhancement of the resistivity so that the power factor is also increased upon crystallization.
The thermopower enhancement is related to energy dependent scattering (energy filtering) of the charge carriers at the nanograin interfaces.
The maximum enhancement was observed for the nearly stoichiometric CrSi$_{\rm 2}$ composites, where the scattering contribution is more than four times larger than the intrinsic thermopower of the amorphous compound.

\section{Acknowledgments}
This work in a part was supported by the Ministry of Education and Science of the Russian Federation in the framework of Increase Competitiveness Program of NUST “MISiS” and by grant Ministry of Education and Science of the Russian Federation RFMEFI58415X0013-14.584.21.0013. SVN gratefully acknowledges financial support by Stipendium of the President of the Russian Federation SP-543.2012.1.

\bibliography{thermoelectrics,nanocomp}

\end{document}